\documentclass[nofootinbib,twocolumn]{revtex4}
\usepackage{graphicx}
\usepackage{amssymb}
\usepackage{amsmath}
\usepackage{bm}
\usepackage{hyperref}

\begin{document}

\title{Structure formation in a $\Lambda$ {\it viscous} CDM universe}

\author{Hermano~Velten}\email{velten@pq.cnpq.br}\author{Thiago~R.~P.~Caram\^es}\email{trpcarames@gmail.com}\author{J\'ulio~C.~Fabris}\email{fabris@pq.cnpq.br}\author{Luciano~Casarini}\email{casarini.astro@gmail.com}
\affiliation{Universidade Federal do Esp\'{\i}rito Santo (UFES), 29075-910, Vit\'oria, ES - Brazil}
\author{Ronaldo~C.~Batista}\email{rbatista@ect.ufrn.br}
\affiliation{Escola de Ci\^encias e Tecnologia, Universidade Federal do Rio Grande
do Norte, Caixa Postal 1524, 59072-970, Natal, Rio Grande do Norte,
Brazil.}

\begin{abstract}
The possibility of dark matter being a dissipative component represents an option for the standard view where cold dark matter (CDM) particles behave on large scales as an ideal fluid. By including a physical mechanism to the dark matter description like viscosity we construct a more realistic model for the universe. Also, the known small scale pathologies of the standard CDM model either disappear or become less severe. We study clustering properties of a $\Lambda$CDM-like model in which dark matter is described as a bulk viscous fluid. The linear power spectrum, the nonlinear spherical ``top hat'' collapse and the mass functions are presented. We use the analysis with such structure formation tools in order to place an upper bound on the magnitude of the dark matter's viscosity. 

\end{abstract}

\maketitle

\section{Introduction}

Dark matter is one of the building blocks of modern cosmology. The prevailing view states that it corresponds to about $25 \%$ of the current matter-energy content  being composed by an unknown form of heavy ($O(m_p) \sim$ GeV) particles which interact very weakly with the remaining cosmic components. Such particles are probably cold relics which induce a large scale background expansion of the form of a cold gas with vanishing kinetic pressure ($p=0$). This description forms the so called Cold Dark Matter (CDM) scenario. 

Although the success in describing the most different astronomical data sets, the full understanding of structure formation process still possess missing concepts. For example, the small scale problems which appear from pure CDM simulations seem to indicate that actual observed large scale structures are not so clumpy as the theoretically simulated CDM agglomeration pattern. 

An alternative proposal like the Warm Dark Matter (WDM) scenario, based on fact that dark matter is composed by a lighter particle with mass of order $\sim$ keV, introduces a cut-off on the clustering power at small scales \cite{wdm}. This happens as a consequence of the larger free streaming length of WDM particles during the radiation dominated epoch. Such process ``erases'' small scale structures more efficiently than CDM and could in principle represent a successful dark matter scenario. However, the confirmation of this hypothesis depends on the positive detection of $m\sim$ keV particles.

In this paper we study an alternative physical mechanism which can mimic in part the WDM success concerning structure formation but still relying on the hypothesis of CDM particles. We add a dissipative term to the description of the CDM stress-energy tensor.

From the possible dissipative mechanisms which can be incorporated to the dark matter description, only the bulk viscosity \cite{Eckart, Chap} remains compatible with the assumption of large scale homogeneity and isotropy. The other processes like shear and heat conduction are directional mechanisms and decay as the universe expands.

Bulk viscous fluids can originate an accelerated model even without dark energy. This happens since this mechanism induces an effective negative pressure to the fluid. Therefore, cosmological scenarios where the bulk viscous fluid coexists only with baryons and radiation (the so called unified models) have been widely studied \cite{viscous1, winfried, VeltenHz, MB}. However, the amount of viscosity needed in the unified scenario in order to accelerate the late time background expansion damages the structure formation
process \cite{Barrow, dominik}. The spherical top-hat collapse was also previously used to confirm such drawbacks faced by a unification scenario based on a bulk viscous fluid \cite{carames}. Therefore, a viable viscous cosmology appears when one uses a usual dark energy component like the cosmological constant to guide the late time expansion whereas only the dark matter fluid is seen as a dissipative fluid.

The proposal of a bulk viscous dark matter component which coexist with a cosmological constant $\Lambda$ was introduced in \cite{dominikVelten2012, velten2014}. We call it $\Lambda$ viscous CDM model ($\Lambda$vCDM). 

If one keeps the same structure like the standard $\Lambda$CDM model, i.e., a dark energy contribution $\sim 70\%$ of the today's critical density then the dark matter's viscosity can not be relevant for the background. In some sense, for a viable cosmological expansion, the larger the value of the dark matter's viscosity, the smaller must be the value of the dark energy density. Therefore, the background expansion depends weakly on the magnitude of the dark matter viscosity. On the other hand, the structure formation process is very sensitive to the viscous mechanism present in the dark matter. The existence of typical dark matter halos in which galaxy formation takes place imposes strong constraints on the magnitude of the dark matter viscosity \cite{velten}. 

In this contribution we explore further constraints of the magnitude of the dark matter viscosity by studying other structure formation techniques. We study the nonlinear evolution of viscous dark matter fluctuations using the spherical collapse model. This result will also enable us to calculate the mass functions of the model. Rather than finding best fit values, the philosophy of our analysis is to set a maximum values for the magnitude of the viscous. Of course, in the case of zero viscosity, our model recovers the $\Lambda$CDM.

We start presenting the background expansion in the next section. It will followed by the perturbative study in sections III (the top-hat collapse) and IV (mass functions). We conclude in the final section.

\section{Cosmological background dynamics and bulk viscosity}

We are interested in flat FLRW background expansions of the type

\begin{equation}
H^2(z)=H^2_0\left[ \Omega_{b0}(1+z)^3+\Omega_{dm}(z) +\Omega_{\Lambda}\right],
\end{equation}
where $H_0$ is the Hubble constant today. We have defined the fractionary density parameters $\Omega_{i}=\rho_{i}/\rho_{crt0}$ for the components $i= b,dm,\Lambda$ (baryons, dark matter and cosmological constant, respectively) where $\rho_{crt0}=3H^2_0/8\pi G$ is the today's critical density.

For the $\Lambda$vCDM model we will identify $\Omega_{dm}$ as as viscous component $\Omega_{v}$. Using the Eckart formalism for dissipative fluids \cite{Eckart} we can assume the existence of a viscous dark matter pressure $P_v$ given by the sum of a kinetic contribution ($p_k$) with the viscous one ($\Pi$), 
\begin{equation}\label{pressure}
P_v=p_k+\Pi= p_k -\xi u^{\gamma}_{; \gamma}
\end{equation}
where $u^{\gamma}$ is the $4-$velocity of the fluid and $\xi$ is the positive coefficient of bulk viscosity. For this parameter it is usual to adopt the following functional form
\begin{equation}
\xi=\xi_0\left(\frac{\rho_v}{\rho_{v0}}\right)^{\nu},
\end{equation}
where $\xi_0$ and $\nu$ correspond to an arbitrary constants and $\rho_v$ is the density of the bulk viscous fluid. As usual, the today's values are denoted by the subscript $0$. The values assumed by these constant parameters are in principle supposed to be positive for thermodynamical reasons \cite{Chap}. 

For a pure viscous dark matter component we assume $p_k=0$ but having in mind that $\left|\Pi\right|\neq 0$ should be very small. In the FLRW metric the bulk viscous pressure reduces to $\Pi =-3H\xi$. The dependence of the pressure $\Pi$ on the expansion factor $H$ imposes severe difficulties to the search for analytical solutions for this model when fluids other than the viscous one, e.g., baryons, are present into the dynamics. For that reason this work will be focused in obtaining only numerical results for the viscous model even at the background level.

The resulting dynamics of the $\Lambda$vCDM model is given by
\begin{equation}
H^2_{\Lambda vCDM}(z)=H^2_0\left[\Omega_{b0}(1+z)^3+\Omega_{v}(z)+\Omega_{\Lambda}\right],
\end{equation}
where the evolution of $\Omega_v$ is therefore obtained from the numerical solution of 
\begin{equation}
(1+z)\frac{d\Omega_v}{dz}-3\Omega_v+\tilde{\xi}\left(\frac{\Omega_v}{\Omega_{v0}}\right)^{\nu}\left[\Omega_{b0}(1+z)^3 +\Omega_{v}+\Omega_{\Lambda}\right]^{1/2}=0.
\end{equation}
The initial condition $\Omega_v(z=0)=\Omega_{v0}$ will be fixed to the same value as $\Omega_{dm0}=0.27$. We also fix $\Omega_{b0}=0.05$.
 
We also have defined the dimensionless parameter 
\begin{equation}
\tilde{\xi}=\frac{24\pi G \xi_0}{H_0}\left(\frac{3 H^2_0}{8\pi G}\right)^{\nu}.
\end{equation}
This definition preserves the relevance of the viscosity. The quantity $\xi_{0}$ carries the dimension $Pa \cdot s$ which is helpful in comparing the viscosity values obtained in our cosmological considerations with other typical values found in nature.

The value $\tilde{\xi}$ is the central aim of this model. Of course, the $\Lambda$CDM model is recovered by setting $\tilde{\xi}=0$, i.e., a pressureless fluid leading to the well know expansion,
\begin{equation}
H^2_{\Lambda CDM}(z)=H^2_0\left[(\Omega_{b0}+\Omega_{dm0})(1+z)^3+\Omega_{\Lambda}\right].
\end{equation}

One may wonder what is the largest value $\tilde{\xi}$ can admit. Of course, for the background expansion $\tilde{\xi}$ plays the same role as $\Omega_{\Lambda}$, i.e., it accelerates the expansion. In general, values of order $\tilde{\xi} < 10^{-1}$ are irrelevant for the homogeneous and isotropic background.
The structure formation process is quite sensitive to values $\tilde{\xi}<10^{-4}$. 
In Ref. \cite{dominikVelten2012} the linear growth of dark matter perturbations was studied. The primordial amplitude of the perturbations were fixed with help of the CAMB code \cite{CAMB}. The perturbation growth of viscous dark matter halos is scale dependent since it carries a $k^2$ term in the Hubble friction part of the equation. Then, the perturbations associated with different scales were allowed to evolve following the differential equation for the viscous linear perturbations. The smaller the scale, the larger the growth suppression. The linear perturbations associated with the dwarf galaxies reach the nonlinear stage only if $\tilde{\xi}\leq 10^{-11}$. Indeed, this is a quite conservative result and must be further studied. On the other hand, the existence of galaxy cluster halos demands $\tilde{\xi}\leq 10^{-6}$. 

We shall verify in the next sections how strong are the constraints coming from the nonlinear top-hat collapse and the analysis of the mass functions.

\section{The nonlinear top-hat collapse}\label{Sec:eq}

Following Refs. \cite{carames, Rui, Abramo:2007iu,  Abramo2} and references therein we study the evolution of an overdense spherical region collapsing in an expanding universe. We firstly obtain quantities and equations which are valid for general fluids. 

For the collapsing region one can write
\begin{eqnarray}
\vec{v}_c= \vec{u}_0 + \vec{v}_p, \\
\rho_c=\rho\left(1+\delta\right) , \\
p_c=p + \delta p.
\end{eqnarray}
The resulting balance between the background expansion and the peculiar motion originates the velocity of the collapsed region $\vec{v}_c$. Hence, the effective expansion rate of the collapsed region can be written as
\begin{equation}
h=H+\frac{\theta}{3a},
\end{equation}
where $\theta=\vec{\nabla} \cdot \vec{v}_p$ and $ \vec{v}_p$ is the peculiar velocity field.

Energy conservation is also required for the collapsing region. Therefore, each component $i$ obeys a separate equation of the type
\begin{equation}\label{eqdeltaorig}
\dot{\delta_i}=-3H(c^2_{eff_i}-w_i)\delta_i-\left[1+w_i+(1+c^2_{eff_i})\delta_i\right]\frac{\theta}{a}
\end{equation}
where the energy density contrast is defined as
\begin{equation}\label{deltadef}
\delta_i = \left(\frac{\delta \rho}{\rho}\right)_i,
\end{equation}
and the effective speed of sound is computed following $c^2_{eff_i} = (\delta p / \delta \rho)_i$.

The dynamical evolution of the homogeneous spherical region will be governed by the Raychaudhuri equation
\begin{equation}\label{eqthetaorig}
\dot{\theta}+H\theta+\frac{\theta^2}{3a}=-4\pi Ga \sum_i (\delta\rho_i + 3\delta p_i)\ .
\end{equation}

In the next subsections we present the specific equations governing the collapse in both $\Lambda$CDM and $\Lambda$vCDM universes.

\subsection{The $\Lambda$CDM model}

Both the baryonic and the dark matter component are assumed to be pressureless fluids. Therefore, we can write down

\begin{equation} \label{blambda}
\dot{\delta}_b=-\left(1+\delta_b\right)\frac{\theta}{a}, 
\end{equation}
\begin{equation} 
\dot{\delta}_{dm}=-\left(1+\delta_{dm}\right)\frac{\theta}{a},
\end{equation}
\begin{equation} \label{thetalambda}
\dot{\theta}+H\theta+\frac{\theta^2}{3a}=-4\pi Ga (\rho_b\delta_b+\rho_{dm}\delta_{dm})\ .
\end{equation}

Since baryons and dark matter obey to similar equations one expects that baryonic matter tracks the dark matter potential wells after the decoupling.

\subsection{The $\Lambda$vCDM model}

Let us now consider the collapse of dissipative dark matter. Note that pressure of the viscous dark matter fluid is dominated by the nonadiabatic contribution $\Pi$, i.e., $p= p_k+\Pi\rightarrow\Pi $

In the Newtonian context the viscous pressure is
\begin{equation}
\Pi=-\xi \vec{\nabla}_r \cdot \vec{u_0},
\end{equation}
therefore its fluctuations read
\begin{equation}
 \delta \Pi=-3H\delta\xi-\xi\frac{\theta}{a}-\delta\xi\frac{\theta}{a}.
\end{equation}

This result leads to a general form for the speed of sound given by  
\begin{eqnarray}
\label{sound}
c^2_{eff_v}&=&\frac{\delta \Pi}{\delta\rho_v}=\frac{-\frac{\xi \theta}{a}-3H\delta\xi-\delta\xi\frac{\theta}{a}}{\delta \rho_v}\nonumber\\&=&\frac{w_v}{\delta_v}\left[\left(\frac{\theta}{3Ha}+1\right)\left(1+\delta_v\right)^{\nu}-1\right],
\end{eqnarray}
where $\omega_{v} \equiv \Pi/\rho_v$ is viscous equation of state parameter. Note that $c^2_{eff_v}$ depends on the divergent of the perturbed velocity field which is a very particular feature of bulk viscous cosmologies.

In order to solve the nonlinear evolution assuming homogeneous fluctuations, we have to check whether pressure gradients are present. Focusing on the viscous pressure, we have  
\begin{equation}
\vec{\nabla} \delta p= \vec{\nabla} \delta \Pi= -3H \vec{\nabla} (\delta \xi)-\xi \frac{\vec{\nabla} \theta}{a}-\vec{\nabla}(\delta\xi\frac{\theta}{a}).
\end{equation}
Assuming $\nu=0$ we have $\delta \xi =0,$ hence 
\begin{equation}
\vec{\nabla} \delta \Pi=-\xi \frac{\vec{\nabla} \theta}{a}.
\end{equation}

Since at high redshifts the viscous dark matter behaves effectively as
pressureless matter, a homogenous profile will not be distorted until
the viscosity becomes important, hence $\vec{\nabla}\delta \Pi$ will be
negligible initially. Moreover, in the most interesting cases $\xi$ is
very small. Therefore we can neglect the remaining part of the pressure
gradient and assume the usual relation $\theta = \theta(t)$ also for the
viscous dark matter model. Under these conditions the viscous pressure perturbation is simply $\delta \Pi=-\xi \frac{\theta}{a}$. From the above considerations we will study only the case $\nu=0$.

Thus considering the introduction of the viscous pressure, the dynamical equations for $\delta_i$ and $\theta$ are modified as follows 

\begin{equation}\label{bv}
\dot{\delta}_b = -\left(1+\delta_b\right)\frac{\theta}{a},
\end{equation}

\begin{eqnarray}\label{dmv}
\dot{\delta}_v=-3H\left(c^2_{eff_v}-w_v\right)\delta_v\nonumber\\-\left[1+w_v+(1+c^2_{eff_v})\delta_v\right]\frac{\theta}{a},
\end{eqnarray}

\begin{equation}\label{thetav}
\dot{\theta}+H\theta+\frac{\theta^2}{3a}=-4\pi Ga \left[\rho_b \delta_b  + \rho_v\delta_v \left(1+3c^2_{eff_v}\right)\right]\ .
\end{equation}

It is important to observe that in these equations we have $c^2_{eff_v}=w_v \theta /3Ha \delta _v$.

\subsection{Results: $\Lambda$vCDM {\it versus} $\Lambda$CDM}

Now we compare the clustering patterns of the $\Lambda$CDM case, by solving equations (\ref{blambda})-(\ref{thetalambda}), 
 and the $\Lambda$vCDM model where we solve the set of equation (\ref{bv})-(\ref{thetav}). We show the results for the growth of the nonlinear dark matter density contrast in Fig. 1, the evolution of the nonlinear baryonic density contrast in Fig. 2 and the expansion rate of the collapsed region in Fig. 3. In both Figs. the red line represents the $\Lambda$CDM. The dashed lines correspond the the viscous dark matter case assuming two different values for the dimensionless viscosity magnitude $\tilde{\xi}$.

We also compute the linear overdensity extrapolated to the collapse time $\delta_c$. This is a fundamental parameter in computation of mass functions in the next section, which is determined by the expression
\begin{equation}
\delta_{c}(z_c)=\delta^l_{v}(z_c),
\end{equation}
where $\delta^l_{v}$ is the linearly evolved viscous density contrast. Its initial conditions are such that the corresponding nonlinearly evolved $\delta_v$ has vertical asymptote at $z_c$, i.e., $lim_{z\rightarrow z^+_c} \delta_{v}(z)= \infty$. 

In Fig.~4 we plot the evolution of $\delta_c$ for the Einstein-de-Sitter universe with $\delta_c=1.686$ (solid horizontal line); for the $\Lambda$CDM model (red line) and for the $\Lambda$vCDM model with values $\tilde{\xi}=0.1$ and $\tilde{\xi}=0.05$ (dashed lines - the caption in the figure identifies the curves).  As we can see, it is clear that the smaller the viscosity parameter, $\tilde{\xi}$, the closer $\delta_c$ is to $\Lambda$CDM values.

\begin{figure}\label{fig1}
\begin{center}
\includegraphics[width=0.43\textwidth]{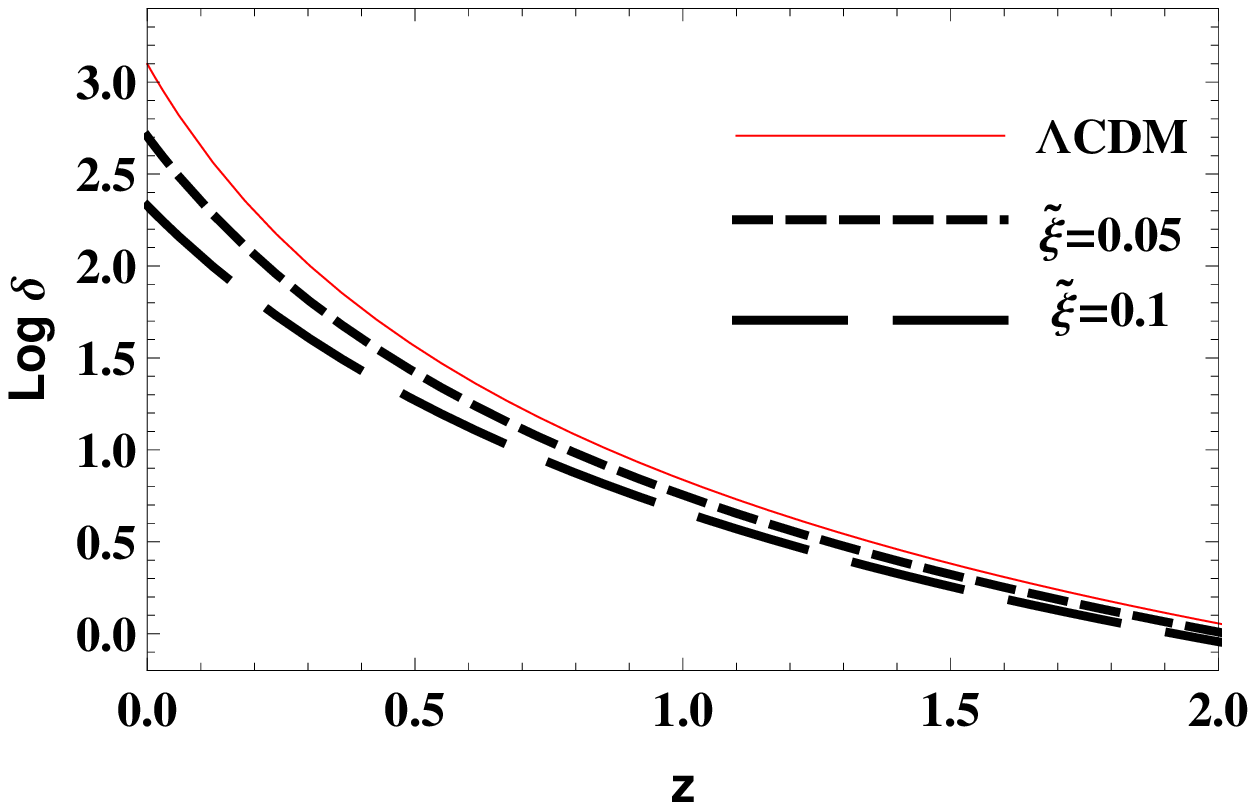}
\includegraphics[width=0.43\textwidth]{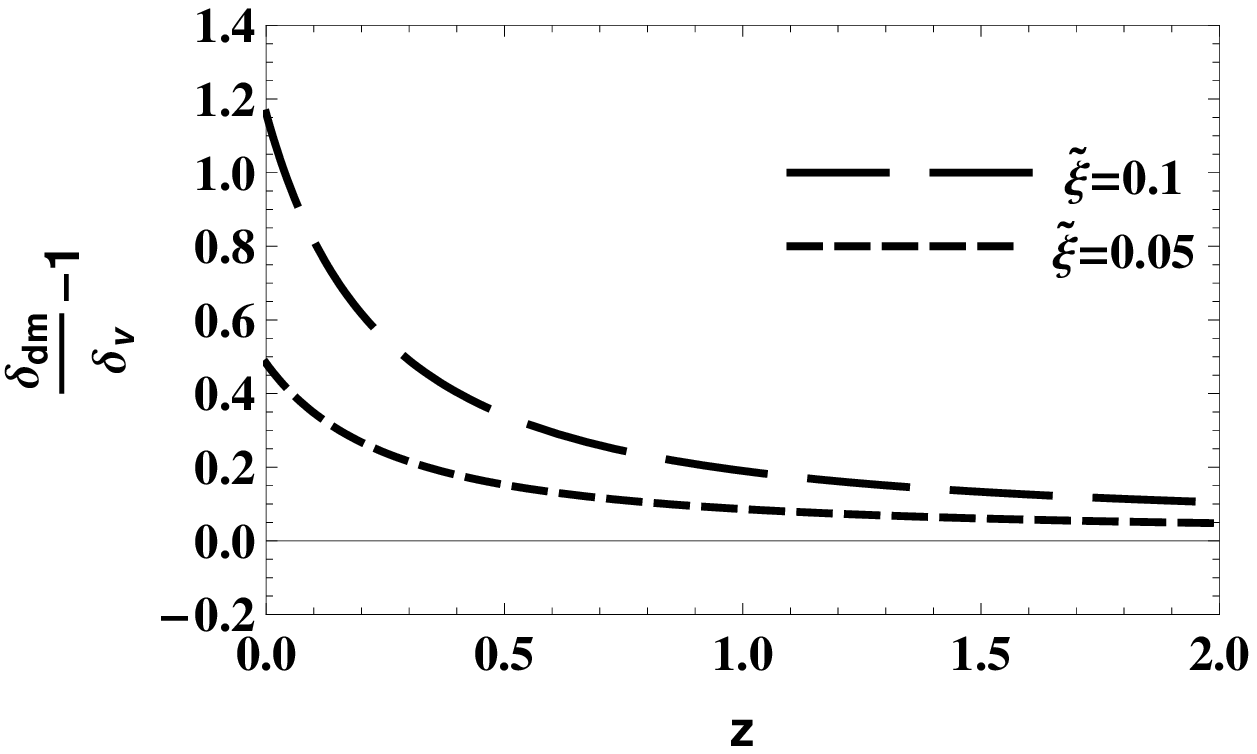}
\caption{Top panel: Dark matter perturbation growth as function of the redshift. Assuming $\Omega_{dm0}=\Omega_{v}=0.27$ with, from top to bottom, $\tilde{\xi}=0.1$ and $0.05$ fixing $\nu=0$. Bottom panel: Relative growth with respect to the $\Lambda$CDM case, i.e., $\frac{\delta_{dm}}{\delta_v} - 1$.}
\end{center}
\end{figure}

\begin{figure}\label{fig2}
\begin{center}
\includegraphics[width=0.43\textwidth]{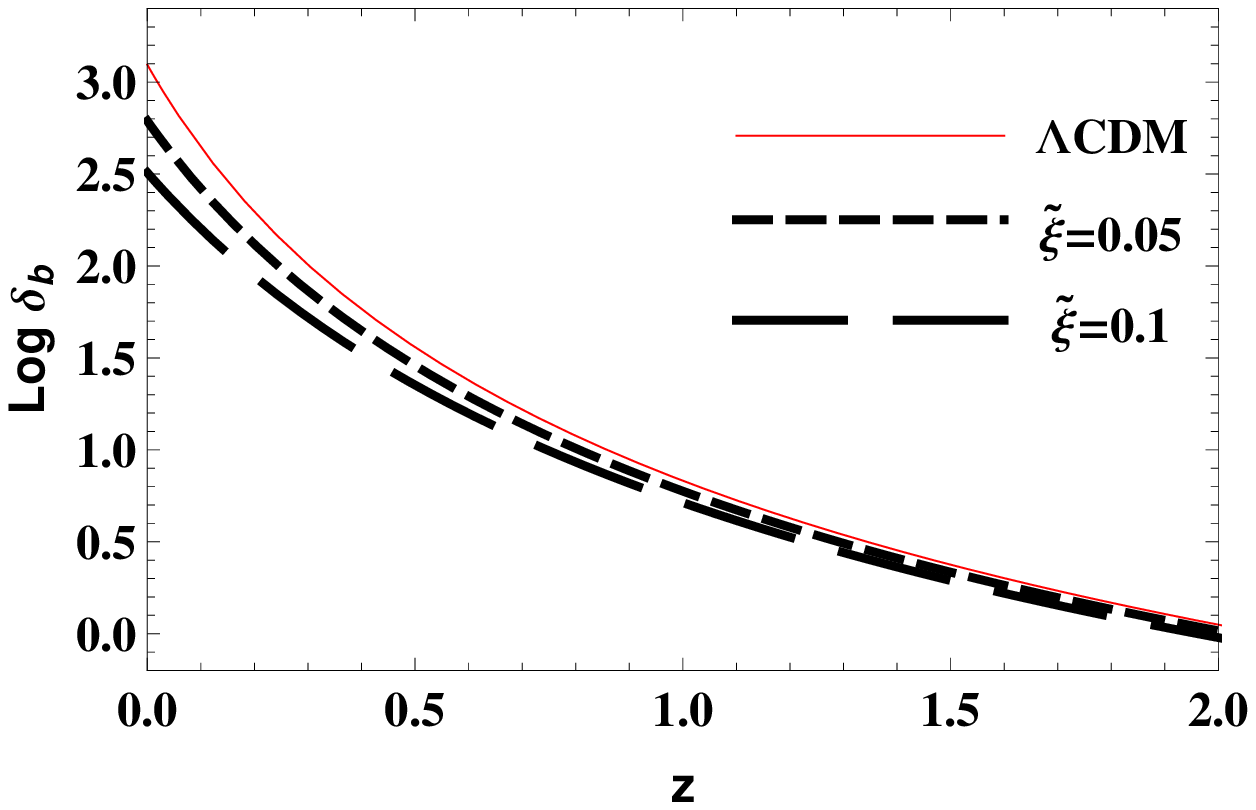}
\includegraphics[width=0.43\textwidth]{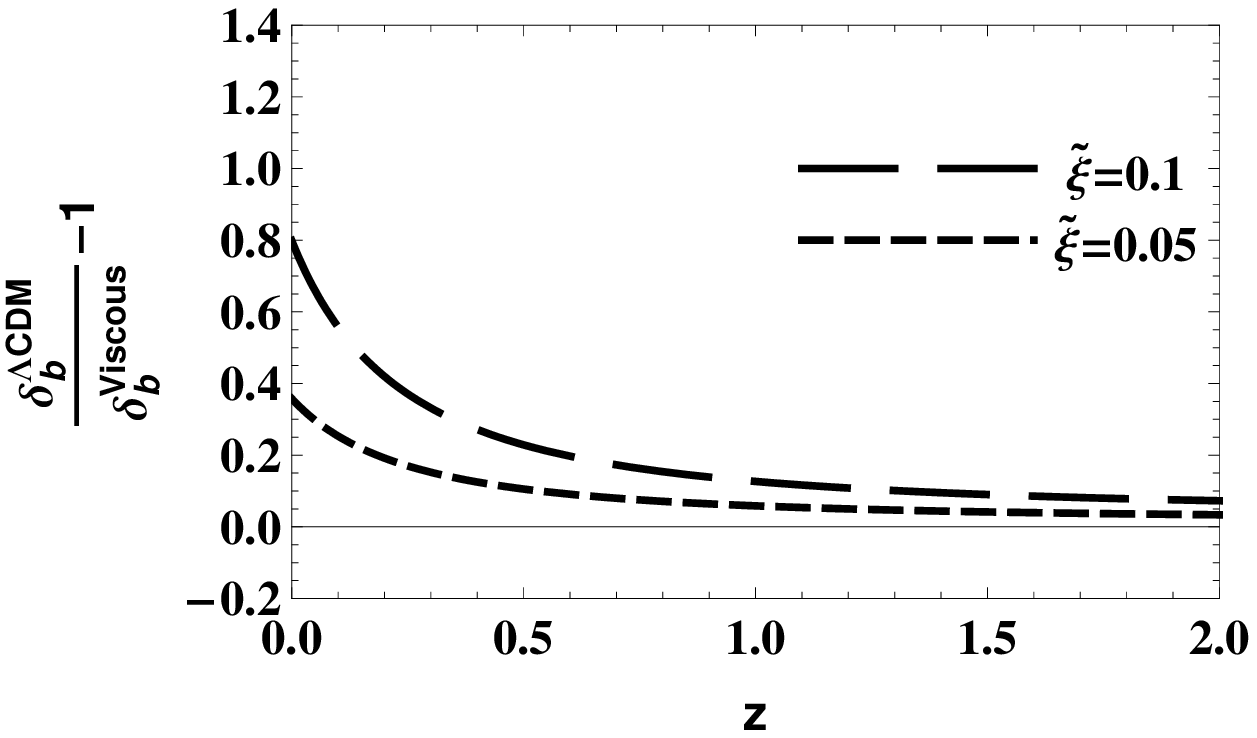}
\caption{Top panel: Baryonic perturbation growth as function of the redshift. $\Omega_{dm0}=\Omega_{v}=0.27$ with, from top to bottom, $\tilde{\xi}=0.1$ and $0.05$ fixing $\nu=0$. Bottom panel: Relative growth with respect to the $\Lambda$CDM case, i.e., $\frac{\delta^{\Lambda CDM}_{b}}{\delta^{\Lambda vCDm}_b} - 1$. }
\end{center}
\end{figure}

\begin{figure}\label{fig3}
\begin{center}
\includegraphics[width=0.43\textwidth]{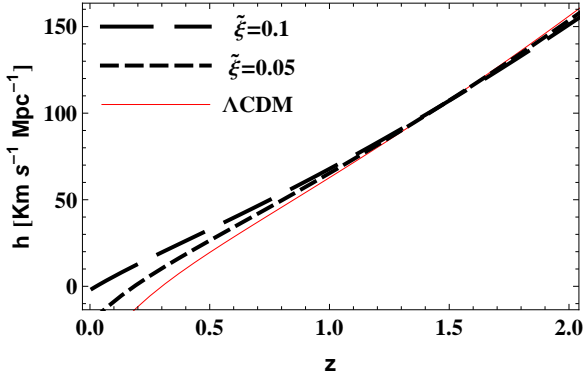}
\caption{Expansion of the collapsed region. $\Omega_{dm0}=\Omega_{v}=0.27$ with, from top to bottom, $\tilde{\xi}=0.1$ and $0.05$ fixing $\nu=0$.  }
\end{center}
\end{figure}

\begin{figure}\label{fig4}
\begin{center}
\includegraphics[width=0.43\textwidth]{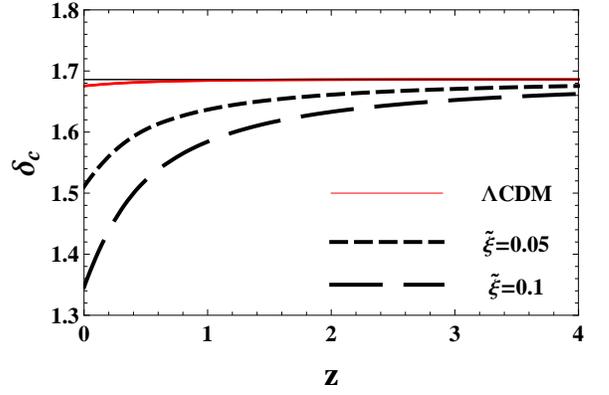}
\caption{Value of $\delta_c$. $\Omega_{dm0}=\Omega_{v}=0.27$ with, from top to bottom, $\tilde{\xi}=0.1$ and $0.05$ fixing $\nu=0$.  }
\end{center}
\end{figure}

\section{Linear Matter Power Spectrum and the Mass function}

As long as the density contrast obeys to $\delta \ll 1$ the perturbation is said to belong to the linear regime. Therefore, the previous analysis concerning the spherical collapse is obviously part of the nonlinear approach. In fact, it provides a general view for the dynamics of the collapse. We can go beyond this and calculate the mass functions which give the abundance of structures as a function of their masses. The mass function is a tool capable to connect the large scale clustering with the number of virialized objects. Its form depends on both the $\delta_c$ values as well as the linear matter power spectrum.

In order to obtain the linear matter power spectrum we follow the approach used in \cite{winfried}. We will be most interested in the effects viscosity causes to the amplitude of density perturbations.
 
In general, for scales larger than the horizon and when pressure effects are important, the evolution of the matter density should be studied using the relativistic perturbation theory. Following \cite{winfried} we write down a general line element for scalar perturbations
\begin{eqnarray}
ds^2= -(1+2\phi)dt^2+2a^2F_{,\alpha}dtdx^{\alpha}\\ \nonumber
+a^2\left[(1-2\psi)\delta_{\alpha\beta}+2E_{,\alpha \beta}\right]dx^{\alpha}dx^{\beta}, 
\end{eqnarray}
where the pertubed 4-velocity is described by $v^0=v_0=-\phi$ and

\begin{equation}
a^2 v^{\mu}+a^2 F_{,\mu}=v_{\mu}\equiv V_{,\mu}
\end{equation} 
which defines the velocity potential $V$. By making the choice $V=0$ we set the comoving gauge. It is also important to introduce $\Xi\equiv a^2(\dot{E}-F)$ which makes the combination $V+\Xi$ gauge-invariant. Indeed, in relativistic perturbation theory it is convenient to describe the dynamics in terms of gauge-invariant quantities representing perturbations on comoving hypersurfaces.

By combining the energy balance with the Raychaudhuri equation, eliminating $V$ and $\Xi$ and transforming the resulting equation to the Fourier ($k-$)space we obtain the second order equation for the linear density contrast $\delta$ (see \cite{winfried} for details)

\begin{equation}\label{lineardelta}
\delta^{\prime \prime}_{v} +f_v(a)\delta^{\prime}_{v}+g_v(a)\delta_v=0,
\end{equation}
where a prime denotes a derivative with respect to $a$ and the coefficients $f_v(a)$ and $g_v(a)$ are
\begin{equation}
f_v(a)=\frac{1}{a}\left[\frac{3}{2}-6\frac{p}{\rho}+3\nu\frac{p}{\rho}-\frac{1}{3}\frac{p}{\rho+p}\frac{k^2}{H^2a^2}\right]
\end{equation}
and
\begin{eqnarray}
g_v(a)=-\frac{1}{a^2}\left[\frac{3}{2}+\frac{15}{2}\frac{p}{\rho}-\frac{9}{2}\frac{p^2}{\rho^2}-9\nu\frac{p}{\rho}\right] \\
+\left[\left(\frac{1}{p+\rho}\frac{p^2}{\rho}+\nu\frac{p}{\rho}\right)\frac{k^2}{H^2a^4}\right],
\end{eqnarray}
respectively. 

At early times, i.e., when $a \ll 1$ equation (\ref{lineardelta}) takes the asymptotic form
\begin{equation}
\delta^{\prime \prime}_{v} +\frac{3}{2a}\delta^{\prime}_{v}-\frac{3}{2a^2}\delta_v=0,
\end{equation}
which has the same solutions as the pure CDM case. The non-adiabatic contributions to the dark matter are sub-dominant on all scales. Then in the past both models are indistinguishable. This will allow us to calculate the viscous matter power spectrum by adopting the same initial conditions as in the standard cosmology.

In Fig.~5 we plot the today's linear matter power spectrum $P(k,z=0)$ for various values of the viscosity magnitude $\tilde{\xi}$. The today's spectrum is calculated by solving the equation (\ref{lineardelta}), evolving the initial transfer function provided by the CAMB code \cite{CAMB}, so evaluating 
\begin{equation}
P(k)=\left|\delta(k,z=0)\right|^{2}.
\end{equation} 
We have assumed for all models the cosmological parameter values $A_s=2.23$ x $10^{-9}$, $n_s=0.96$, $k_{pivot}=0.05$ Mpc$^{-1}$, $h=0.67$ and $\Omega_{\Lambda}=0.69$.

The suppression at small scales is evident for $\tilde{\xi}$ values in the range $10^{-5} > \tilde{\xi} > 10^{-7}$. Even though the considered values of $\tilde{\xi}$ are very small, the fluctuation amplitudes at $8 h^{-1}$ Mpc differ considerably from the standard model. 

\begin{figure}\label{fig5}
\begin{center}
\includegraphics[width=0.51\textwidth]{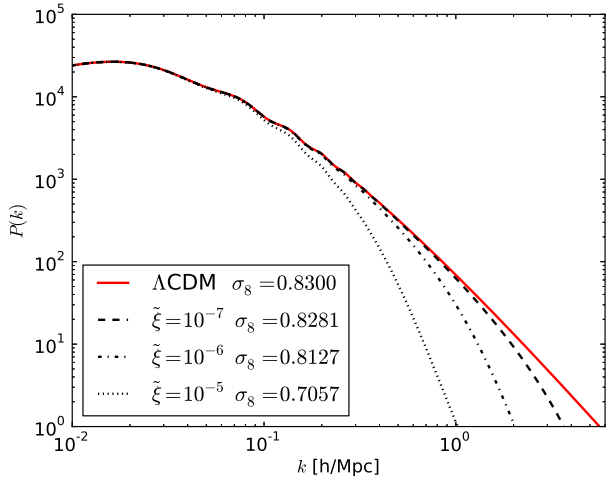}
\caption{Linear matter power spectrum at $z=0$, for viscosity parameter in the range $10^{-5} > \tilde{\xi} > 10^{-7}$, the relative computed values of $\sigma_8$ are also showed; red solid line represents the 
$\Lambda$CDM reference.}
\end{center}
\end{figure}

Since we have derived the linear matter power spectrum we can calculate the mass function. We can build the number count of clustered objects integrating the universal recipe of mass function $n(m,z)$ for a mass $m$ and a redshift $z$ :
\begin{equation}
\eta\,f(\eta)= m^2\,{n(m,z)\over \rho}\,
{{\rm d}\,\ln m\over {\rm d}\,\ln \eta},
\,\,
\eta=\delta_c/\sigma(m).
\label{fnunm}
\end{equation}
Supposing that the fluctuations at early times have spherical shapes, we can compute the mass variance $\sigma(m)$ as function of scale $r \propto m^{1/3}$, and $\delta_c(z)$ is the linear-theory density contrast at the time of spherical collapse at redshift $z$. Also, $\rho$ is the background density, and the distribution of the first crossings  $\eta f(\eta)$  is provided for ellipsoidal collapse by \cite{st}:
\begin{equation}
\eta\,f(\eta) = A\,\Bigl[1 + (a\,\eta)^{-p}\Bigr]\ 
\left({a\,\eta\over 2}\right)^{1/2}\,{{\rm e}^{-a \eta/2}\over\sqrt{\pi}} ,
\label{giffit}
\end{equation}
where $a=0.707$, $p=0.3$ and $A=0.3222$.
Following the results showed in Fig.~1, we can consider the linear density contrast of viscous model indistinguishable from the $\Lambda$CDM one at $z=0$, in the range of $10^{-5} > \tilde{\xi} > 10^{-7}$. Of course, the same happens for the $\delta_c$ value and therefore we can keep the value $\delta_c=1.69$ for viscous models with very low $\tilde{\xi}$ values.

Figure~6 compares the cumulative mass function $N(>M)$ at $z=0$ for these values of the viscosity parameters. Coherently with Fig.~5, the suppression due to the viscosity is substantial also for tiny values, and it is more evident at smaller scales.
As the matter power of the viscous models approaches or overlaps the $\Lambda$CDM one, the mass function coincides also at large scales. 

\begin{figure}\label{fig6}
\begin{center}
\includegraphics[width=0.51\textwidth]{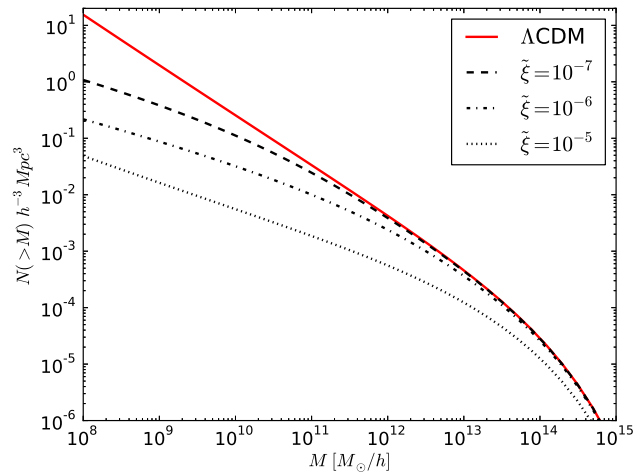}
\caption{Sheth \& Thormen predictions \cite{st} at $z=0$, for viscosity parameter in the range $10^{-5} > \tilde{\xi} > 10^{-7}$, assuming linear density contrast $\delta_c=1.69$; red solid line shows the $\Lambda$CDM reference.}
\end{center}
\end{figure}

\section{Conclusions}

We have investigated an extension of the standard cosmological scenario in which the cold dark matter is assumed to possess a dissipative property in the form of bulk viscosity (the $\Lambda$vCDM model). 

Our main goal here was to place an upper bound on the magnitude of the dark matter viscosity $\tilde{\xi}$ by using both the linear and the nonlinear cosmological perturbation theory. The spherical collapse study performed in section III is not so sensitive to the $\tilde{\xi}$ value when compared to the power spectrum or the mass functions analysis. However, this analysis is crucial to show that the value of the critical density $\delta_c$, a fundamental parameter for constructing the mass functions, is very similar to the value assumed in the $\Lambda$CDM universe for low viscosities $\tilde{\xi}<10^{-5}$. 

The linear power spectrum shown in Fig.~5 reveals that if dark matter possess a viscosity of order $\tilde{\xi}=10^{-5}$ then one observes a strong growth suppression at small scales. For the sake of comparison, the effect of a $\tilde{\xi}=10^{-6}$ dark matter viscosity on the power spectrum is very similar to a warm dark matter scenario with particle masses of order $\sim$ keV \cite{wdm}. 

Finally, when the mass functions are computed we find out that a value $\tilde{\xi}=10^{-7}$ reduces in one order or magnitude the abundance of number counts with masses of order $M\sim 10^{9}M_{\odot}$. 

In Ref.\cite{dominikVelten2012} the linear growth of some typical astrophysical scales was studied. By requiring that dwarf galactic scales (once such scales are observed) should reach the non linear stage of evolution, i.e., $\delta=1$, the conclusion was that values $\tilde{\xi}>10^{-11}$ would erase these small perturbations. However, many aspects of the structure formation process like for example the merger history which can propel the agglomeration process were not taken into account and therefore the bound $\tilde{\xi}\lesssim10^{-11}$ can be overestimated. 
 
From our results it seems fair to conclude that values $\tilde{\xi}\approx10^{-8}$ will lead to almost the same predictions as the standard $\Lambda$CDM model but with a slightly suppressed growth of the smallest cosmological structures in the universe. 

The dark matter viscous pressure leads to a damping on the growth of structures which is proportional to the dark matter's viscosity magnitude. It is therefore expected that for a given value of the bulk viscosity the viscous dark matter proposal will not show the same clustering patterns as the standard cosmology and the so called small scale problems of the typical CDM paradigm would not show up. In particular, our results indicate that the viscous dark matter candidate can be seen as a potential solution for the missing satellites problem \cite{satellites}. The final answer to the issue will come from future full N-Body simulations of viscous dark matter particles. Here, we have not performed such simulations, however we did an useful intermediary step towards this problem.

\textbf{Acknowledgement}:  We thank CNPq (Brazil) and FAPES (Brazil) for partial financial support. R.C.B thanks FAPERN for the financial support.

\end{document}